\documentclass[12pt,preprint]{aastex}

\usepackage{graphicx}
\usepackage{multirow}

\slugcomment{To appear in PASP}

\shorttitle{ROVOR}
\shortauthors{Moody et al.}

\begin{document}

\title{Remote Observatory for Variable Object Research (ROVOR)}

\author{J. W. Moody, B. Boizelle, K. Bates, B. Little, T. McCombs, J. Nelson, C. Pace\altaffilmark{1}, R. L. Pearson III\altaffilmark{2}, J. Harrison\altaffilmark{3}, P. J. Brown\altaffilmark{4}}
\affil{Physics and Astronomy Department, Brigham Young University,
    N283 ESC, Provo, UT, USA 84602}
\email{jmoody@byu.edu}

\and

\author{J. Barnes}
\affil{Salt Lake Community College Physics Dept., Salt Lake City, UT, USA 84070}

\altaffiltext{1}{Current address: Indiana University, Dept. of Astronomy, Swain Hall West 319, Bloomington, IN, USA 47405-7105}
\altaffiltext{2}{Current address: University of Denver, Dept. of Physics \& Astronomy and The Observatories, 2112 East Wesley Avenue, Denver, CO, USA 80208}
\altaffiltext{3}{Current address: Northern Arizona Unversity, Dept. of Physics \& Astronomy, PO Box 6010, Flagstaff, AZ, USA 86011-6010}
\altaffiltext{4}{Current address: Mitchell Institute for Fundamental Physics \& Astronomy, Department of Physics \& Astronomy, Texas A\&M University, College Station, TX, USA 77843-4242}

\begin{abstract}
Observatories constructed solely for photometric monitoring make it possible to understand the temporal nature of objects over time scales that historically have been difficult to achieve. We report on one such observatory, the Remote Observatory for Variable Object Research (ROVOR) which was constructed to enable both long-term and rapid cadence observations of brighter objects. ROVOR is an 0.4m optical telescope located in central Utah and commissioned for scientific observations in 2008. Principle research has been monitoring blazars, x-ray binaries, AGN, and an occasional gamma-ray burst afterglow. We describe the observatory, the control system, and its unique roof.
\end{abstract}


\section{Introduction}

\label{sec:intro}

In the past half-century observational astronomy has benefited from technical innovations which have steadily led to larger aperture telescopes and more sensitive detectors. These advancements have revolutionized the science, directly or indirectly bringing about most of the significant discoveries of this period. They have also driven an evolution toward greater scale, complexity and expense causing the economics of building and operating large observatories to become an increasingly significant factor in astronomical research. To maintain high productivity, it is incumbent that institutions minimize costs whenever possible. One way to reduce costs is to use the minimum equipment that is adequate for the research.

In the past two decades public interest has led to the development of research-quality mounts and smaller-aperture telescopes that are relatively inexpensive.  These systems are appropriate for some types of imaging research---in particular monitoring the optical variability of brighter objects. As a result, temporal monitoring is now achievable at more sophisticated levels than before.

Temporal monitoring has a rich history.  The American Association of Variable Star Observers (AAVSO) has archived more than 21 million observations as of the end of 2011. These data give fundamental information about stellar properties such as mass, radius, luminosity, temperature, internal and external structure, composition, and evolution. Understanding these aspects of cepheid variable stars, for example, has allowed them to play a major part in determining distances to the nearest galaxies, and from that, the expansion rate and age of the Universe. Accretion disks surrounding supermassive black holes are of great interest in physics and cosmology. How their energy varies in time is one of the only direct measures of the environment and structure of these unresolved objects.

Even so, the long baselines and especially the frequency of observation required for insightful data can be demanding and often unrewarding as data are compiled through uneventful epochs. Because of this, optical variability data are commonly limited to cadences that are less than optimum. For example, there have been several significant long-term radio monitoring campaigns of blazars and quasars \citep[see][and references therein]{Bach06} but only the prototypical objects such as 3C273, 3C66a, AO 0235$+$164, S5 0716+714, 3C 454.3, BL Lac, Mrk 501, Mrk 421, etc. are monitored with any dedicated regularity. The vast majority of blazars and quasars remain unmonitored and knowledge of their behavior and structure advances slower than would otherwise be the case.

However, this is changing. Many impressive surveys have been or are being conducted using less expensive, smaller aperture telescopes. Among these are Near-Earth Object searches \citep{LINEAR, NEAT, Drake09}, extrasolar planet surveys \citep[e.g.][]{HATNet, SuperWASP, Nutz08, DEMON, KELT}, imaging optical afterglows of gamma-ray bursts \citep[e.g.][]{PROMPT, Pi}, variable stars and transients \citep[e.g.][]{AAVSOnet,APASS,ROTSE,PTF}\footnote{A significant survey of brighter stars is the Bright Star Monitor. See http://www.aavso.org/bsm.} and supernovae \citep[see][and references therein]{Howell09}.  Surveys for one type of variable object usually find other types as well, such as comets and novae, illustrating how fertile temporal monitoring is for discovery.

Pan--Starrs \citep{PanStarrs}, SkyMapper \citep{SkyMap} and the Large-Scale Synoptic Telescope \citep[LSST]{LSST} are and will be enormous boons to temporal studies of all kinds. Even so, since they cover large areas of the sky to relatively deep magnitude limits, they are constrained to slower cadences and fainter objects than can be observed with smaller systems that are dedicated to a few objects. Even after the advent of the LSST, smaller aperture telescopes will be necessary to monitor the brighter, prototypical objects---those which often provide great insight into the behavior of the object class.

We report here on an observatory dedicated to variable studies named the Remote Observatory for Variable Object Research or ROVOR. ROVOR was conceived as a means of exploring the science capabilities of a small remote telescope dedicated to monitoring the variability of brighter objects.  The intent is to use it as a springboard toward establishing a cluster of similar systems. But it has been productive by itself, providing data for blazar monitoring campaigns \citep{Abdo11a, Abdo11b}, gamma-ray burst afterglow imaging \citep[e.g.][]{Pearson}, and several on-going AGN and x-ray binary monitoring projects \citep{Boizelle}.

From the outset ROVOR was designed to 1) be inexpensive to build, 2) be easy to maintain, 3) be robust with little down time, 4) be operable with minimally trained labor, and 5) provide significant data. It must be inexpensive since a functional system that is costly to run might tax limited resources to the point of making it unproductive. It must be robust and easy to maintain since we have no-one within 100 miles to service the telescope. It is truly remote with visits for maintenance and trouble-shooting only as required. Fortunately private companies have developed and marketed smaller but excellent systems for reasonable costs, completing much of the first three tasks. (It is fortunate that the general public loves and supports astronomical imaging!) The last two points are more an obligation of higher education than industry and we have addressed them ourselves.

Most of the labor for building and operating ROVOR has come from undergraduate upperclassmen. Their academic placement between beginning and graduate students characterizes the operational placement of ROVOR. It is to produce data of scientific significance while also educating astronomy majors in the basics of astronomical research.

In \S 2 we present the basic data on the ROVOR facility including the telescope, detector and building.  In \S 3 we review the control of the facility, in \S 4 we discuss observational capabilities and in \S 5 we discuss lessons learned and future plans.

\section{The Facility}

\begin{figure}
   \epsscale{.80}
   \plotone{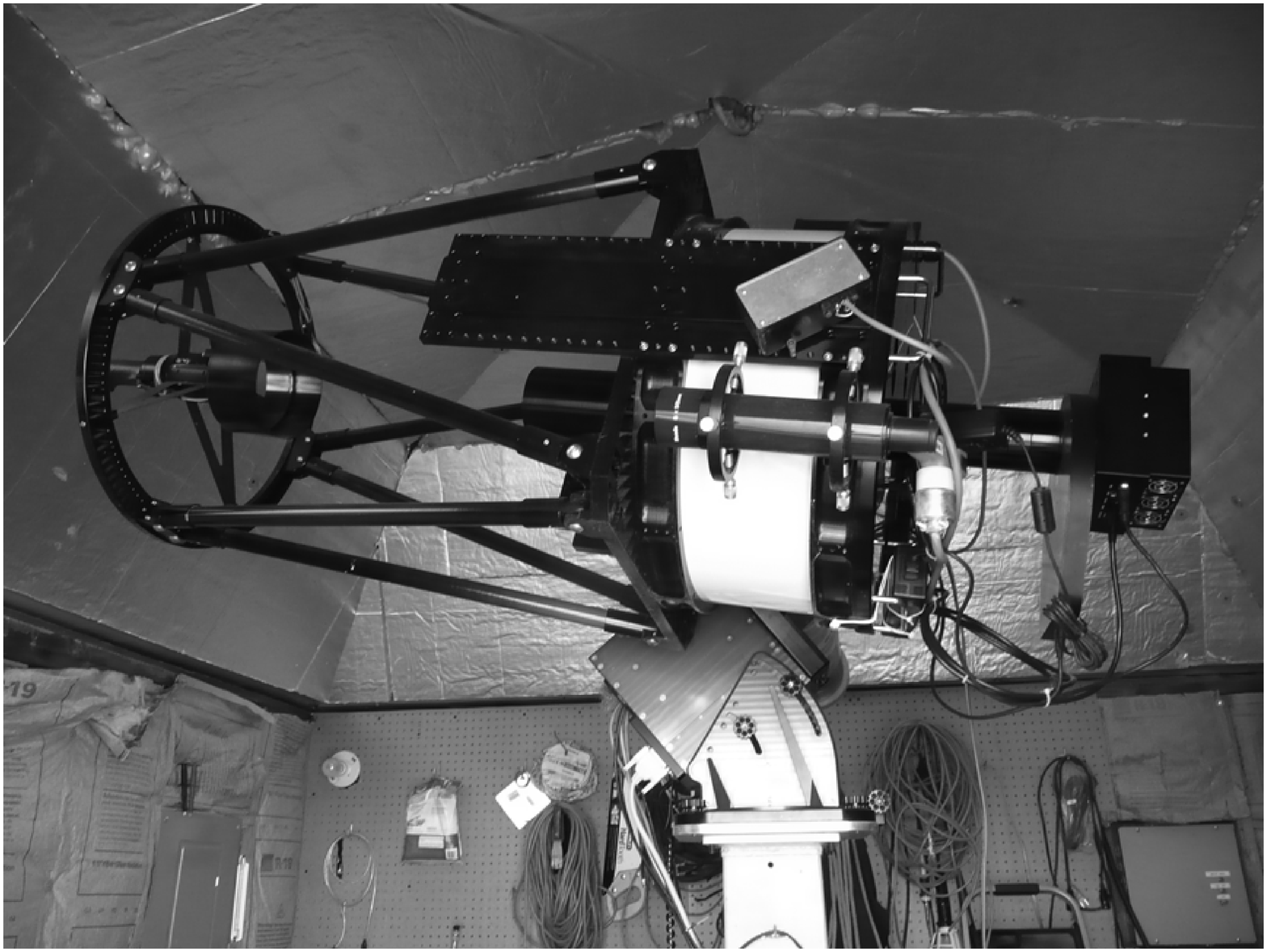}
   \caption[f1]
   { \label{fig:f1}
The telescope in park postion. The optical tube is a Ritchey-Cretien 0.4m from RC Optical. The mount is a German-equatorial Paramount ME. The FOV is $23.4'$ on a side with a resolution of $1.37 ''$/pixel.}
   \end{figure}

The ROVOR observatory houses a remotely operated 0.4m telescope located in central Utah at 112\arcdeg43$'$01.0$''$W 39\arcdeg27$'$17.1$''$N  and 4579 feet above sea level.  The observatory was initiated when an unfinished 0.6m Autoscope telescope was donated to the BYU Physics and Astronomy Department. An observatory was built to house it and the telescope was installed. When this telescope unfortunately proved to be unusable, it was replaced with an RC Optical tube on a German-equatorial Paramount ME pier, similar to the telescopes and mounts of the PROMPT and MEarth arrays.  The tube is an open truss f/9 with a primary mirror ion milled to 1/30 wave rms (see Fig. \ref{fig:f1}).

The telescope is equipped with an imaging CCD and filter wheel. The filter wheel is a six-position FLI CFW-6-6 equipped with 79mm round parafocal Johnson-Cousins BVRIClear filters. The CCD camera is an FLI ProLine PL003 with a 1024 x 1024 24 \micron~pixel SITe detector. The FOV is $23.4'$ on a side with a resolution of $1.37 ''$/pixel.  Pointing is accurate to better than half an arcminute. There is no guiding and individual exposures are typically limited to less than two minutes to minimize or eliminate trailing.  Within a $45^{\circ}$ zenith angle, the images are rarely trailed more than one arcsecond.

The ROVOR site is located 12 miles west of Delta, Utah at the end of the local power grid and on a valley floor for ease of access. It is within the eastern side of the detector grid of the ``Telescope Array'' international cosmic ray air shower observatory\footnote{See http://www.telescopearray.org/}, one of the few expansive dark areas left in the continental US. Seeing is fairly consistent at approximately $3''$ FWHM.  The wide FOV and $1.37 ''$/pixel resolution are a good match for the conditions and make optimum seeing values less critical. While less-accessible sites with better seeing were available, we chose this site knowing our anticipated FOV and pixel scale were an adequate match to those conditions. This choice illustrates that ROVOR has been built to explore the economically possible more than the absolutely ideal.

\begin{figure}
   \epsscale{.70}
   \plotone{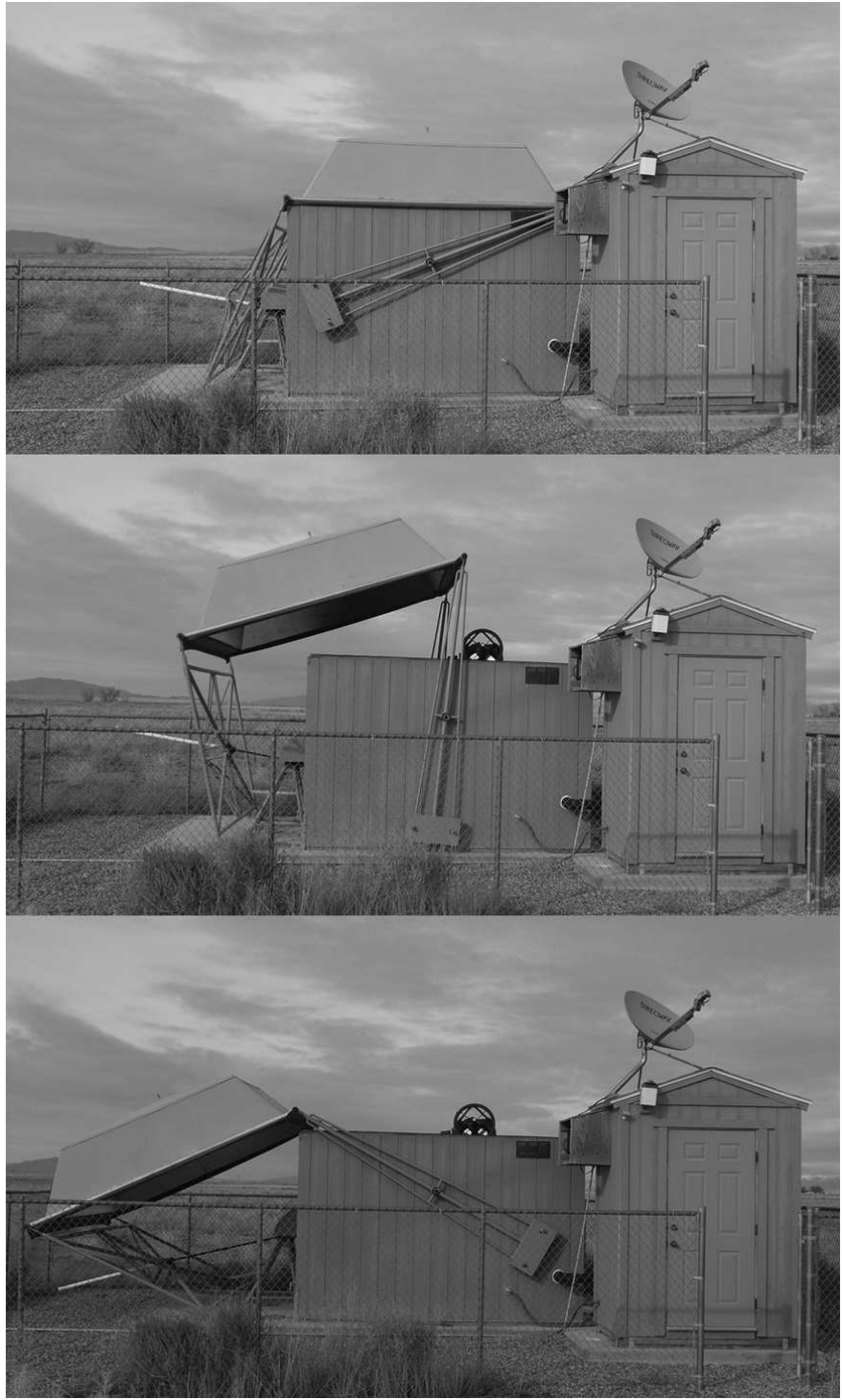}
   \caption[f2]
   { \label{fig:f2}
The ROVOR facility showing the roof closed (top), partially opened (middle), and fully open (bottom). The roof lifts up and over the telescope then lowers to allow the telescope an unobstructed view to within approximately 10 degrees of the western horizon. The electronics shed (called the ``outhouse'') containing the communication and control hardware is on the right.  The roof pull-off mechanism is on the left side.}
   \end{figure}

ROVOR is housed in a simple 10 foot x 10 foot wooden shed, dubbed the ``doghouse''.  The roof pivots off in a manner similar to the lids of the HATNet telescopes (see Fig. \ref{fig:f2}). Called a ``Lifferth'' dome after its designer Wes Lifferth, the roof lifts over the telescope and down to the west side of the building, giving some protection against the prevailing winds. This design has an advantage over a roll-off roof for remote observing because there is no track to keep clean.

\begin{figure}
   \plotone{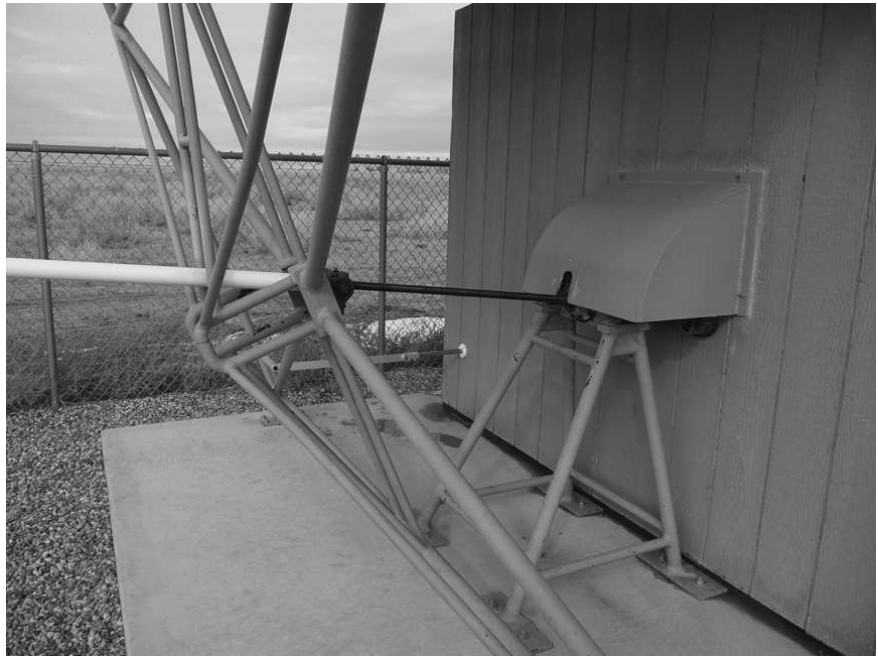}
   \caption[f3]
   { \label{fig:f3}
A close-up of the roof pull-off mechanism. A threaded rod pushing on a gimballed brass knuckle in the middle of a rigid steel frame at the back of the roof pivots the roof up and over the telescope.  A white pvc pipe protects the rod from dust. The roof position sensor is a rod on the north-west corner seen penetrating into the building.}
   \end{figure}

The roof is moved by a 1.5 horsepower electric motor twisting a 6 foot long 1.5 inch diameter steel threaded rod.  At 110V it uses approximately 10 amperes and takes 2 minutes 20 seconds to open or close.  The rod pushes on a gimballed threaded brass sleeve attached to a rigid steel framework at the back of the roof (see Figs. \ref{fig:f2} and \ref{fig:f3}). This awkward-looking mechanical actuator was chosen over alternatives like an hydraulic system since it was simpler with fewer potential failure points.

The telescope can image to the horizon on the east, north, and south and to within 10 degrees of the horizon on the west. A small patch in the southeast, well below normal viewing angles, is obscured by the communication dish. Since the telescope can rise above the building walls, a mercury gravity switch on its side prevents the roof from moving whenever the telescope is not stowed safely out of the way.

The paint scheme is a non-standard brown for the sides and blue for the roof. At the remote location there is a concern for vandalism. The brown sides and blue roof are for camouflage against the horizon and actually do reduce the visibility of the building. Traditional white is of course better for temperature regulation. However, this is not as critical for a removable roof which, when pulled completely off, allows the temperature to rapidly equalize.

Control computers are housed in a small building to the side called the ``outhouse''. This structure also holds an all-sky camera and the communications dish.  The outhouse controls the temperature of the electronics and prevents them from being exposed to the weather when the roof is off.  A weather station on the property perimeter provides constant information on wind speed and direction, humidity, temperature and visibility.

\section{Communication and Control}

The site is controlled by two computers communicating with each other through a LAN router. One computer manages the telescope, roof, and detector while the other manages the weather station, data transfer, security cameras, all-sky camera and communication link.

The data connection is through a HughesNet satellite link.  We chose a satellite link over a land line for convenience and to learn how to successfully deal with its speed and latency issues.  Satellite links are available world-wide and are easy to establish while a telephone link may not always be possible in truly remote locations. In the spirit of this project we felt it prudent to learn now how to live with the more readily available option.

There are two ways to control the instrumentation.  First is by remote terminal with the computers themselves. We use the software suite from Software Bisque of {\it{Orchestrate}}, {\it{CCDSoft}}, and {\it{The Sky}} to control the telescope and detector.  We use our own software written in {\it{LabView}} to open and close the roof.  Satellite latency makes accessing these programs slow, but the speeds are still adequate for setting up a single night's observation. It also allows us to examine the general health of our systems as if we were sitting at the telescope itself.

We define latency as the time delay from when a data request is initiated to when the data are received. The physical distance between the facility and the satellite and especially the data verification overhead and other internal electronic delays within the satellite itself give us a latency that can be as long as 1500ms. This means that a single click of the mouse can take over a second to successfully process. We cannot remove the latency from our system but we can mitigate its inefficiency by using fewer commands to accomplish the same tasks. To do this, we have developed a software package called {\it{CelestialGrid}} to talk to the above programs using multiple instructions embedded into data packets. This way we bypass the command overhead, such as screen updates, that comes with remote terminal programs.
 
To run the telescope using {\it{CelestialGrid}}, the user first assembles an observation sequence from a directory of objects stored within it. {\it{CelestialGrid}} guides them as they put the list together by indicating the hour angle of each observation. When ready, the list is encoded and queued for transmission at a certain time specified by the user. Once observing commences, the information is compressed and sent to the telescope. {\it{CelestialGrid}} runs the telescope in a nearly robotic mode; starting the night's observations at twilight, taking calibration and flat frames, conducting the observations, and parking the telescope when finished.  It also has a feed from the Gamma-ray burst Coordinates Network (GCN).  When interruption is enabled, ROVOR will automatically image any bursts that are above the horizon from receipt of the pointing vector until sunrise.

On average, {\it{CelestialGrid}} reduces by a factor of 16 the number of communications required to accomplish the same command using a remote terminal. It also enables us to script certain common observation sequences and keep them on site so that they do not need to be transmitted via satellite.  We prefer to operate with {\it{CelestialGrid}}, reverting to the remote terminal to trouble-shoot or to run more specialized observing programs that are as yet outside {\it{CelestialGrid}}'s capabilities.

{\it{CelestialGrid}} itself has two separate components: a client side application installed on site and the main server software which can be installed on any machine. The client side mainly receives and executes scripts as they are transmitted to it from the main server.  It also understands the execution sequence of complex operations like taking flat fields in several filters so these can be accomplished via a single "go" command. Both packages were developed in {\it{Oracle Java}} to make the server software platform-independent.  However the client side does require {\it{Microsoft Windows}}. The server software is distributed among ROVOR team members as a {\it{Java Web Start}} application which enables all installations of the server software to automatically update on startup. This feature has been very useful in rapidly deploying updates and fixes.

The client software still uses {\it{Orchestrate}} to communicate with the telescope and camera. {\it{Orchestrate}} has the convenient feature of executing scripts sequentially as they are placed in a folder. After executing the scripts, it outputs them to a second folder with specific comments about the success of each element within the script. {\it{CelestialGrid}} feeds {\it{Orchestrate}} small individual scripts and reads the output results. Because scripts fed to {\it{Orchestrate}} are only for a few individual images on a single object at a time, the observing can be interrupted for sudden transient events like a gamma-ray burst by having the main server simply inject a script into the on-going sequence.

The {\it{CelestialGrid}} server provides a command-line batch utility for archiving images.  Images are uploaded either manually or automatically through ftp at the end of each night and placed on a terabyte hard disk with redundant backup for safety. After data processing, both raw and reduce frames are archived into a {\it{MySQL}} database.

\section{System Capabilities}

We estimate the absolute photometric accuracy of ROVOR from data obtained for a study of the standard stars around TeV blazars \citep[see][]{Pace}. In this research BVRI magnitudes were re-measured for standard stars around the blazars listed in \cite{TeV}. These measurements were standardized with respect to stars from \cite{Landolt09}.

A standard photometric transformation for the $V$ magnitude has the form
\begin{equation}
V = v_o + a_{vo} + a_{v1}(B_o - V_o),
\end{equation}
where $V$ = the standardized $V$ magnitude, $v_o$ = instrumental magnitude, $a_{vo}$ and $a_{v1}$ are the transformation coefficients, and $B_o - V_o$ is the color term. The $B_o - V_o$ color term is calculated via
\begin{equation}
B_o - V_o = a_{b1}(b_0-v_0)+a_{b0}.
\end{equation}
where $b_o - v_o$ is the instrumental color term and $a_{bo}$ and $a_{b1}$ are transformation coefficients,.  Similar equations are used for the other filters. The transformation coefficients for a typical night, in this case August 21, 2009, are given in Table~\ref{tbl-1} columns 1-3. The coefficients themselves are not terribly interesting for our purposes here.  However their uncertainty is on the order of 0.01 magnitudes or less meaning the entire system is fairly stable.

We obtained data on seven standard stars around Mrk 501 multiple times a night through eight independent nights. The stars ranged from 12th to 16th magnitude. The exposure times per filter are given in column 4 of Table 1. The standard deviation of the values obtained for all eight nights is given in column 5. Photometric values for these same stars are published by \cite{FT96}, \cite{Vill98}, and \cite{Dor05}. Column 6 gives the offset between the ROVOR values and the mean of the published values defined as published - ROVOR.

The values of the standard deviations of the offsets range from 0.03 to 0.046 magnitudes.  The offsets themselves range from 0.008 to 0.032 magnitudes.  The standard deviations are all within the errors quoted for our comparison sources which means that our own observational errors are not contributing as much to the uncertainty as the standards are.  So we can conservatively conclude that our typical absolute photometric values are accurate to at least 0.03 magnitudes and are likely better than this.

\begin{figure}
   \plotone{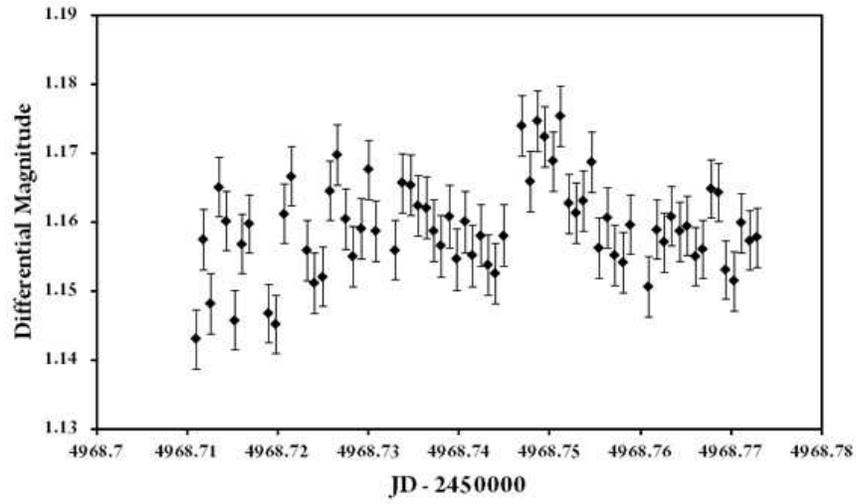}
   \caption[f4]
   { \label{fig:f4}
Photometry of Mrk 501, a 14th magnitude blazar, for 17 May 2009. Each data point is a 60 second exposure in R. The one-sigma error bar for each photometric measurement is less than 0.005 magnitudes. A change in brightness of approximately 0.02 magnitudes at a time of 4968.75 is apparent.}
   \end{figure}

The precision of differential ROVOR photometry can be estimated by a second study on the intranightly variations of Mrk 501 undertaken by \cite{RLPIII}. Fig. \ref{fig:f4} shows photometric data taken on the night of 17 May 2009. The y-axis plots the magnitude difference between Mrk 501, which has a V magnitude of approximately 14, and a standard star of similar magnitude within the FOV. The x axis plots the time throughout the night. The exposures were all 60 seconds and the photometry was through a six arcsecond diameter aperture. The internal error bars are between 0.003 and 0.005 magnitudes. A jump in brightness of 0.02 magnitudes is apparent at a time of 4968.75. The scatter in values early in the night is on this order and is likely intrinsic to Mrk 501.  It is this type and level of variability that ROVOR is designed to monitor.

While the individual errors from 17 May 2009 are better than average, it is still apparent that on a typical clear night ROVOR can confidently detect magnitude variations of 0.01 and on a good night can see changes of half that amount. This precision is commonly accomplished with similarly sized telescopes.

\begin{figure}
   \plotone{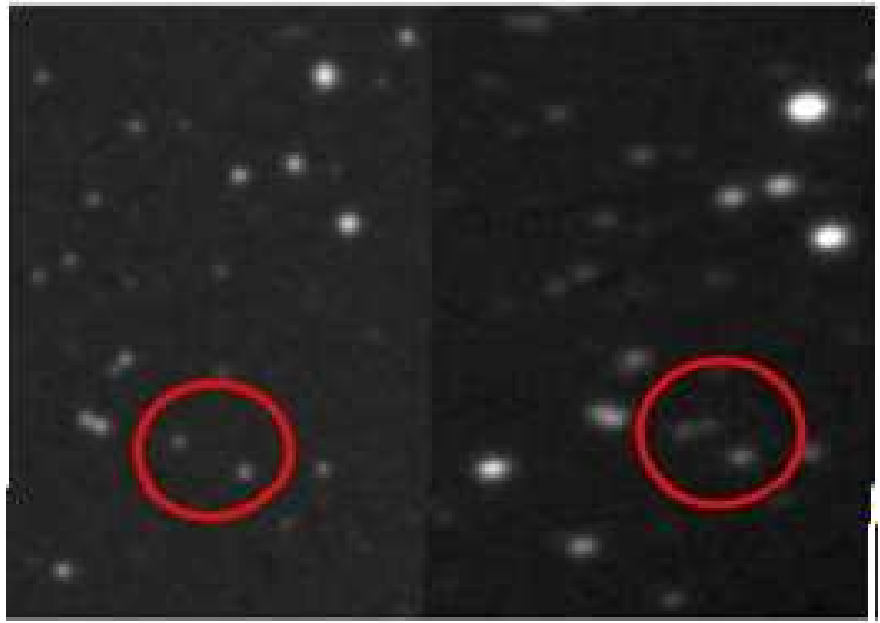}
   \caption[f5]
   { \label{fig:f5}
A before (left) and after (right) image of GRB100418, captured 19 April 2010, 9.75 hours after the initial detection by the SWIFT satellite. The burst image is visible in the circle on the right. The image is from the coaddition of 28 two-minute exposures in R taken through a mean air mass of 2.48.  The magnitude of the burst at this time was 18.2 in R}
   \end{figure}

Finally Fig. \ref{fig:f5} of GRB100418A is included to show the ability to image gamma-ray burst afterglows.  Fig. \ref{fig:f5} is from the addition of 28 two-minute images taken through an R filter on 19 April 2010, 9.75 hours after the burst was detected.  The object was low on the horizon at an airmass of 2.48 which degraded the image quality somewhat.  Even so, the object, at a magnitude of 18.2 in R is clearly visible.  The limiting magnitude for an unfiltered detection is approximately R = 22 in one hour.

\section{Discussion}

The successes of the ROVOR project are many. We routinely obtain differential magnitudes with millimagnitude precision. The focus and tracking are stable and the  sub-arcminute pointing is adequate. We are pleased but not surprised at the depth with which we can probe for gamma-ray burst afterglows. As mentioned in the introduction, it is certain that even in the LSST era, ROVOR-like telescopes will be used to constantly monitor at a higher cadence and/or in specialized filters the brighter, prototypical objects that are saturated in larger aperture surveys.

We have had very few hardware problems with the telescope and none with the CCD. The roof has functioned flawlessly for four years with the only glitches being a bolt that worked loose, dropping the north pivot arm to the ground and a relay that welded closed forcing the roof past the open limit, dumping it onto the ground. Both were easily repaired. And so far there have been no bullet holes in the walls although an enraged bull did bend the threaded rod on the roof pull-off mechanism.  We have yet to invoke our ultimate backup for roof failure; a large tarp and a cell-phone call to a kind local farmer.

The biggest problem has been the communication link which has failed in several novel ways including a fried satellite modem, a computer virus, and a wasp nest in the satellite feed horn. We accept problems of this nature as the ongoing price to be paid but are still looking at how we might simplify and fortify the link further. It is wise to pay a regular monthly visit to check on the hardware, swap out parts before they fail, and regularly update the virus data base.

A significant challenge has been relying on undergraduate labor to get the system working.  Over 20 undergraduate astronomy majors have worked on ROVOR.  As soon as they become proficient, they leave and training begins again. This is a great model for education but a poor model for timely progress.

Future effort will be concentrated on the data analysis pipeline. Currently data are uploaded automatically the morning after observing for analysis at BYU. We are working toward having all analysis take place on site so that only calibrated magnitudes need be transmitted.  We are also redesigning the server portion of {\it{CelestialGrid}} to be a web application that can control several telescopes at once. We are bringing {\it{CelestialGrid}} into full ASCOM compliance and are exploring ways to minimize the computing overhead on the telescope control computer to make real-time computer problems less likely.

\acknowledgments

We are greatly indebted to David Crawford for his inspiration and John Meriwether who first taught JWM what remote observing with modest instruments could accomplish. We thank David Derrick and the BYU college of physical and mathematical sciences for providing the funds to build the ROVOR facility.  We are indebted to Michael Joner and Dan Reichart for several insightful discussions and recommendations. John Ellsworth has provided expert electronic help and Wes Lifferth has taken care of all mechanical needs.  And we thank Theo Berry, Travis Smith and Kevin Styler for being willing to check on the facility, turn off lights, and shut doors for absent-physically and absent-minded professors.

{\it Facilities:} \facility{ROVOR}.

\clearpage

\clearpage

\clearpage

\clearpage

\clearpage

\clearpage

\begin{table}
\caption{Photometric accuracy \label{tbl-1}}
\begin{tabular}{ccccccc}
\tableline\tableline
Filter & Coefficient & Value ($\sigma$) & exposure & scatter & offset \\
 & &(10 Aug 2012) & (sec) & (mag) & (mag) \\
\tableline
\multirow{2}{*}{B} & $a_{b0}$ & ~0.224 (0.026) & \multirow{2}{*}{240} & \multirow{2}{*}{0.046} & \multirow{2}{*}{0.008}\\
  & $a_{b1}$ & ~1.303 (0.005) &  &   \\
\multirow{2}{*}{V} & $a_{v0}$ & 19.931 (0.010) & \multirow{2}{*}{180} & \multirow{2}{*}{0.033} & \multirow{2}{*}{0.009}\\
  & $a_{v1}$ & -0.083 (0.005) &  &   \\
\multirow{2}{*}{R} & $a_{r0}$ & ~0.095 (0.010) & \multirow{2}{*}{120} & \multirow{2}{*}{0.030} & \multirow{2}{*}{0.018}\\
  & $a_{r1}$ & -0.940 (0.004) &  &   \\
\multirow{2}{*}{I} & $a_{i0}$ & 0.334 (0.027) & \multirow{2}{*}{90} & \multirow{2}{*}{0.031} & \multirow{2}{*}{0.032}\\
  & $a_{i1}$ & ~0.864 (0.006) &  &   \\
\tableline
\end{tabular}
\end{table}

\end{document}